# Slow light microfluidics: a proposal

M. Sumetsky

*Aston Institute of Photonic Technologies, Aston University, Birmingham B4 7ET, UK*
*m.sumetsky@aston.ac.uk*

The resonant slow light structures created along a thin-walled optical capillary by nanoscale deformation of its surface can perform comprehensive simultaneous detection and manipulation of microfluidic components. This concept is illustrated with a model of a 0.5 millimeter long 5 nm high triangular bottle resonator created at a 50 micron radius silica capillary containing floating microparticles. The developed theory shows that the microparticle positions can be determined from the bottle resonator spectrum. In addition, the microparticles can be driven and simultaneously positioned at predetermined locations by the localized electromagnetic field created by the optimized superposition of eigenstates of this resonator, thus, exhibiting a multicomponent near field optical tweezers.

Recent developments of microfluidics in engineering, physics, chemistry, biochemistry, nanotechnology, and biotechnology have demonstrated remarkable progress in understanding microfluidic behavior and its applications (see, e.g., [1-4]). The basic goal of microfluidics is the effective extraction of detailed characteristics of microfluids and in-situ manipulation of their ingredients. Optics and photonics provide a powerful means to address this twofold goal by measuring the refractive index, fluorescence, Raman scattering, etc. [3, 4]. In addition, the employment of optical forces enables trapping and manipulation of microfluidic components [5]. The rapidly developing field of research bridging optics and microfluidics is named optofluidics [3, 4]. The intriguing challenge of optofluidics consists in realization of in-situ sensing and manipulation of microfluids on a chip without application of free space beams and bulk optics (such as microscopes and optical beam tweezers, as examples). Ideally, the entire control and manipulation of an optofluidic structure should be performed through the input-output optical waveguides connecting an optofluidic chip to a light source and detector. In spite of significant achievements [3-9], this goal still lacks a comprehensive solution.

This Letter proposes an approach to address the mentioned critical challenge of microfluidics using the recently introduced platform for the fabrication of miniature photonic slow light structures at the surface of an optical fiber. The platform is called Surface Nanoscale Axial Photonics (SNAP) [10-13]. The fabrication precision of SNAP is two orders of magnitude higher (whilst the transmission losses are two orders of magnitude smaller) than for any of the previously reported platforms. The new technology uses resonant light that circulates along the optical fiber surface and slowly propagates along the fiber axis. By varying the fiber radius by only a few nanometres, multifunctional SNAP devices can be created, e.g., complex-shaped bottle resonators including circuits of coupled resonators [11-13] (Fig. 1(a)).

The basic idea of this Letter is to create resonance SNAP structures at the surface of an optical (silica) capillary with a micron-thin wall and use these structures for evanescent sensing and manipulation of microfluids inside the capillary. As proposed by White, Oveys and Fan [14, 15] and developed in numerous publications (see, e.g., references in [4]) the variation of a high Q-factor spectral resonance of a whispering gallery mode (WGM) excited in a *uniform* micron-thin silica capillary by a transverse microfiber or planar waveguide can be used for accurate detecting of the local refractive index change in microfluids. So far, the detection was performed locally and was based on the measurement of variation of a *single resonance*.

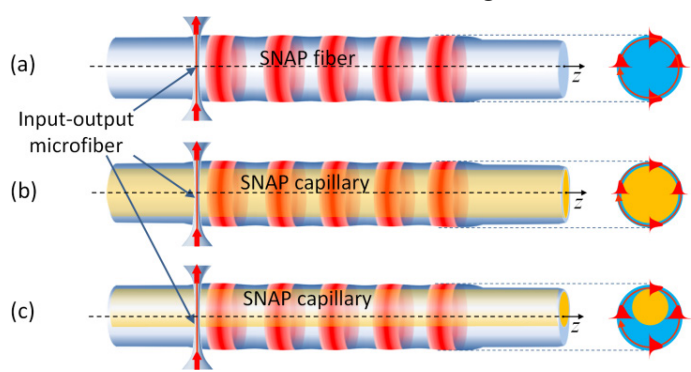

Fig. 1. (a) Illustration of a SNAP structure created at the fiber surface and coupled to the input-output microfiber; (b) The same at the capillary surface; (c) The same at the surface of the fiber with a capillary asymmetrically positioned close to the fiber surface. The right hand side of (a), (b), and (c) illustrate the circulation of light in the device cross-sections and its penetration into the microfluid.

It is conjectured here that the SNAP resonance structures created at the surface of an optical capillary with a wavelength-scale thin wall allow one to extract both *spatial and temporal dependences* of physical and chemical characteristics of liquids in microchannels as well as to *manipulate the microfluidic processes* with evanescent fields. In addition to the axially symmetric uniform capillary having a few micron-thin wall (Fig. 1(b)), one can consider the asymmetrically positioned capillary with a relatively small fraction having the micron wall thickness (Fig. 1(c)). In the latter case, the evanescent field is strongly localized along the azimuthal (as well as the radial) direction and can perform sensing and manipulation of microfluids in a small micron/submicron cross-sectional area adjacent to the thinnest part of the capillary. It is shown below that the spatial and temporal characteristics of microfluids along the length of a specially designed SNAP structure within this cross-sectional area can be extracted from the spectrum of this structure.

As an example, this Letter theoretically demonstrates the detection and manipulation of microparticles floating in a thin highly transparent capillary with a SNAP triangular bottle resonator created along the capillary surface. It is shown that the positions of microparticles can be (i) determined by monitoring the spectrum of a SNAP resonator and (ii) manipulated by variation of the evanescent field of this resonator.

The field distribution of a WGM slowly propagating along the axis $z$ of an optical capillary containing spatially uniform microfluid can be written in the cylindrical coordinates $\mathbf{r} = (\rho, \varphi, z)$ as $E_{m,p}(\mathbf{r}) = e^{im\varphi}\Xi_{m,p}(\rho)\Psi_{m,p}(z)$, where $\Psi_{m,p}(z)$ is defined by the one-dimensional Schrödinger equation [11]

$$\Psi_{zz} + (E(\lambda) - V_0(z) - W_{int}(z,t))\Psi = 0. \quad (1)$$

Here the azimuthal and radial quantum numbers, $m$ and $p$, are omitted for brevity. In Eq. (1), the energy $E(\lambda) = -\kappa^2 \Delta\lambda / \lambda_{res}$ is a linear function of wavelength variation $\Delta\lambda = \lambda - \lambda_{res} - i\gamma$. The potential $V_0(z) = -\kappa^2(\Delta r(z)/r_0 + \Delta n(z)/n_r)$ corresponds to the uniformly-filled capillary. Here $r_0$ is the fiber radius, $\Delta r(z)$ and $\Delta n(z)$ are the fiber radius and index variations, $\kappa = 2^{3/2}\pi n_r / \lambda_{res}$, $\lambda_{res}$ is resonance wavelength, $n_r$ is the refractive index of the fiber, and $\gamma$ determines the attenuation of light in the fiber. Potential $W_{int}(z,t)$ in Eq. (1) is a function of $z$ and *parametric* function of time $t$ which describes the processes in the microfluid adjacent to the internal capillary surface. Generally, these processes are determined by the 3D perturbation of the refractive index, $\Delta n_{int}(\mathbf{r},t)$, caused by the microfluidic spatial and temporal variations. Subsequently, the transmission amplitude $A(\lambda,t)$ measured at the output of the microfiber coupled to the capillary (Fig. 1) [10-13] is used to monitor the microfluidic processes in space and time.

Let us consider micron-size particles floating along the microfluidic capillary. After approaching the capillary surface, a microparticle can be captured by the evanescent field of the SNAP resonator and continue to move along the capillary surface. For the symmetric configuration (Fig. 1(b)), a microparticle will drift in two (axial and azimuthal) dimensions. For the asymmetric configuration (Fig. 1(c)), a microparticle will propagate along the line parallel to the capillary axis following the local maximum of the evanescent field.

Since the axial propagation of light in a SNAP structure is slow [11, 13] the characteristic wavelength of the Schrödinger equation, Eq. (1), is much greater than the radiation wavelength (typically, greater than 10 μm). Therefore, it is assumed here that the size of detected microparticles is much smaller than this characteristic wavelength. Then, the microparticles can be modelled by Eq. (1) with the short-range potentials:

$$W_{int}(z,t) = \sum_{n=1}^{N} D_n(t)\delta(z - z_n(t)) \quad (2)$$

Here $N$ is the number of microparticles, $\delta(x)$ is the delta-function, $z_n(t)$ determines the axial coordinate of microparticle, and $D_n(t)$ determines its detection contrast. The value of $D_n(t)$ is expressed through the 3D refractive index variation $\Delta n^{(n)}(\mathbf{r} - \mathbf{r}_n(t))$ caused by the microparticle $n$:

$$D_n(t) = -\frac{\kappa^2 \eta_{m,p}}{2\pi r_0 n_r} \int d\mathbf{r} \Xi_{m,p}^2(\rho) \Delta n^{(n)}(\mathbf{r} - \mathbf{r}_n(t)), \quad (3)$$

where $\eta_{m,p} = \left(\int d\rho \Xi_{m,p}^2(\rho)\right)^{-1}$ is the normalization factor of $\Xi_{m,p}^2(\rho)$. In the derivation of this equation, it was assumed that the characteristic size of the evanescent field along the azimuthal direction (Fig. 1(c)) is much greater than the microparticle size. This is always true for the axially symmetric capillary (Fig. 1(b)). From Eq. (3), the contrast $D_n$ rapidly grows with decreasing of the capillary wall thickness. However, thinning of the wall causes the decrease of the resonance Q factor. As opposed to the axially symmetric capillary (Fig. 1(b)), the micron-thin area in the asymmetric capillary (Fig. 1(c)) can be relatively small, and, thus, less affect the degradation of the Q-factor.

The Green's function of Eq. (1) with short-range potentials determined by Eq. (2) can be analytically expressed through the Green's function of Eq. (1) with potential $V_0(z)$ in the absence of microparticles [16]. Thus, the transmission amplitude of a capillary with floating microparticles, $A(\lambda,t)$, can be determined as well. In this case, the solution of the inverse problem, which expressed $D_n(t)$ and $z_n(t)$ through $A(\lambda,t)$ is reduced to the solution of a system of $2N$ algebraic equations at discrete $\lambda_n^j$, $j = 1, 2$.

The high Q-factor of the SNAP resonance structures enables highly accurate microfluidic sensing by monitoring the shifts $\Delta\lambda_s(t)$ of resonances (here $s$ is the resonance number). For a relatively small variation of refractive indices $\Delta n^{(n)}(\mathbf{r},t)$, the shift $\Delta\lambda_s(t)$ is determined by the perturbation theory:

$$\Delta\lambda_s(t) = -\lambda_{res}\kappa^{-2}\sum_{n=0}^{N} D_n(t) |\Psi^{(s)}(z_n(t))|^2. \quad (4)$$

Here $z_0(t) \equiv z_0$ is the coordinate of the microfiber and $D_0(t) \equiv D_0$ is the resonator/microfiber coupling parameter [11], which are usually independent of time $t$ and can be determined experimentally [11, 12]. Functions $\Psi^{(s)}(z)$ in Eq. (4) are the normalized eigenfunctions of the resonator determined from Eq. (1).

As an example, consider a microparticle moving in the evanescent field of a triangle SNAP bottle resonator created at the surface of a silica capillary with radius $r_0 = 50\,\mu m$ and micron-thin wall (Fig. 2(a)). The potential inside the triangular resonator is

$$V_0(z) = -\kappa^2(\Delta r_0 / r_0)(1 - z/L). \quad (5)$$

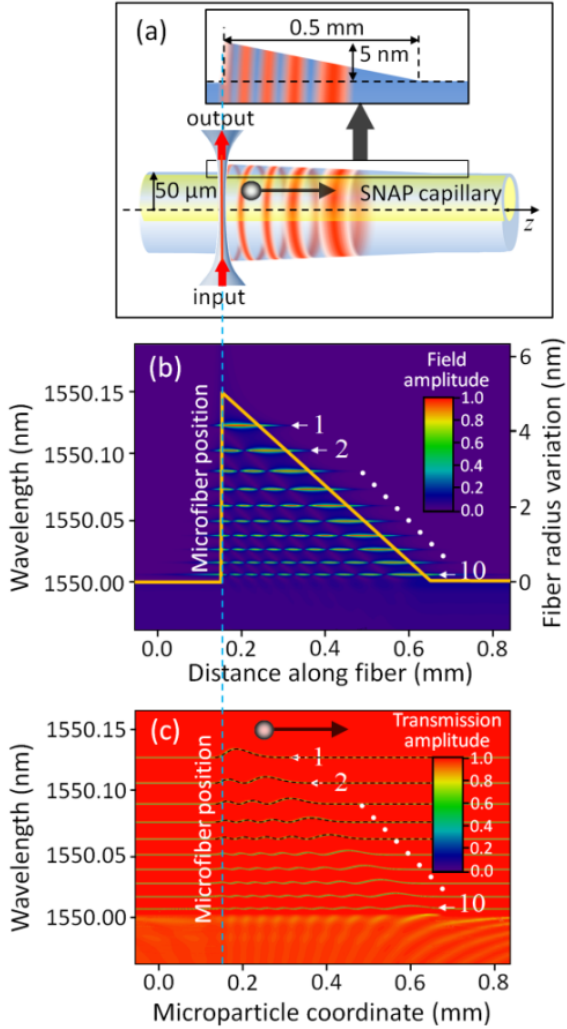

Fig. 2. (a) – A capillary with the introduced triangular SNAP bottle resonator coupled to the transverse input-output microfiber. Inset: fiber radius variation magnified along the radial direction. (b) – 2D plot of the resonant evanescent field calculated at the capillary surface as a function of axial coordinate and wavelength. Arrows with numbers indicate the wavelength eigenvalues. (c) – The numerically calculated resonant transmission spectrum of the SNAP structure shown in Fig. 2(a) as a function of the axial coordinate of the microparticle moving along the internal capillary surface and wavelength. Dashed lines are the first five resonance trajectories calculated from Eq. (7).

The effective height of the triangle is $\Delta r_0 = 5$ nm and its length is $L = 0.5$ mm (inset in Fig. 2(a)). The spectrum of the microcapillary sensor is measured using the microfiber positioned at the left edge of the resonator as shown in Fig. 2(a). The intrinsic Q-factor of resonator is set to $10^6$, the value measured in [15]. The microfiber-resonator coupling parameters are assumed to be equal to those measured in [12]. The resonance wavelength is set to $\lambda_{res} = 1.55$ μm. Then, for a silica microcapillary with $n_r = 1.45$ we have $\kappa = 8.31\,\mu\text{m}^{-1}$. Fig. 2(b) shows the 2D plot of the resonant evanescent field as a function of axial coordinate $z$ and wavelength $\lambda$. It is seen that the resonator possesses 10 axial eigenmodes. While the first eigenmode is localized in the vicinity of the left edge, the tenth mode is distributed along the whole length of the resonator. The axial dimension of modes grows linearly with their number.

The microparticle is modelled by the potential in Eq. (1) corresponding to the effective refractive index variation with the Gaussian shape,

$$W_{int}(z,t) = -\kappa^2 (\Delta n_0 / n_r) \exp[-(z - z_1(t))^2 / w^2]. \quad (6)$$

Here the refractive index maximum $\Delta n_0$ is estimated from Eq. (3), $\Delta n_0 = \eta_{m,p}(2\pi^{3/2} r_0 w)^{-1} \int d\mathbf{r}\, \Xi_{m,p}^2(\rho) \Delta n^{(1)}(\mathbf{r})$, and, as noted above, strongly depends on the width of the capillary wall. The preferable configuration for the detection of microparticles is the asymmetric capillary (Fig. 1(c)) since it permits the enhancement of the microparticle-resonator coupling by thinning of a relatively small fraction of the capillary with minimum reduction of the resonator Q-factor. In the numerical simulations shown in Fig. 2(c), the characteristic index variation is set to $\Delta n_0 = 0.0005$ and $w = 0.6$ μm corresponds to the axial FWHM of the microparticle equal to 1 μm. This figure shows the numerically calculated surface plot of transmission amplitude as a function of microparticle position $z_1$ and wavelength $\lambda$. The trajectories of sharp spectral dips in this plot determine the variation of eigenvalues of the triangular resonator as a function of the microparticle position. The normalized eigenmodes of triangular resonator can be approximately expressed through the Airy functions:

$$\Psi^{(s)}(z) = A_s \left(\frac{\kappa^2}{r_0}\right)^{1/6} \text{Ai}\left(-\left(\frac{\kappa^2}{r_0}\right)^{1/3} \left[z - \left[\frac{3\pi}{2}\left(s - \frac{1}{4}\right)\right]^{2/3}\right]\right) \quad (7)$$

where $A_s$ is the normalization factor, $A_1 = 1.426$, $A_2 = 1.245$, $A_3 = 1.156, ...$ Comparison of rescaled $\Psi^{(s)}(z)$ calculated from Eq. (7) for $s = 1, 2, ...5$ (dashed lines) with the trajectories of resonances at the surface plot of Fig. 2(c) demonstrates remarkably good accuracy of Eq. (7). Nonlinear Eqs. (4) and (7) can be used to determine the microparticle position $z_1(t)$ and its contrast $D_1(t)$ as functions of time. Generally, these two unknowns can be determined from two equations of Eq. (4) corresponding to $s = s_1$ and $s = s_2$ provided that eigenfunctions with numbers $s_1$ and $s_2$ vary fast enough near the microparticle position. For example, one can use $\Psi^{(1)}(z)$ and $\Psi^{(2)}(z)$ near the left hand side of the resonator. However, the inclusion of functions $\Psi^{(s)}(z)$ with larger $s$ is necessary to determine the microparticle position closer to the right hand side where $\Psi^{(1)}(z)$ and $\Psi^{(2)}(z)$ vanish. The plot in Fig. 2(c) can be used for detection of microparticles with an effective refractive index $\Delta n_0$ different from the value 0.0005 assumed above. To this end, the amplitude of resonance trajectories in this figure should be rescaled by the factor $\Delta n_0 / 0.0005$.

The resonant evanescent field of the SNAP structure created at the capillary surface can be used to control the microfluidic processes and, in particular, to manipulate the floating microparticles which are attracted to the capillary surface by this field. The strongly localized driving field can be created using a multi-parametric tunable filter positioned at the input of the microfiber. The filter launches light into the SNAP resonator at $M$ eigenvalues $\lambda_1, \lambda_2 ..., \lambda_M$ with the required partial powers and phases chosen so that the superposition of the excited modes is localized at the predetermined axial position. Fig. 3 demonstrates this for the triangular bottle resonator considered. In this figure the linear combinations of ten complex-values axial distributions of the evanescent field calculated near the ten resonant wavelengths (Fig. 2(b)) are optimized to arrive at the single-maximum trapping field which can drive a microparticle within the axial dimensions of the resonator (Fig. 3(a)) and the double-maxima trapping field which can independently and simultaneously drive two microparticles within the axial dimensions of the resonator (Fig. 3(b)). In both cases, only real (field intensity) coefficients were optimized.

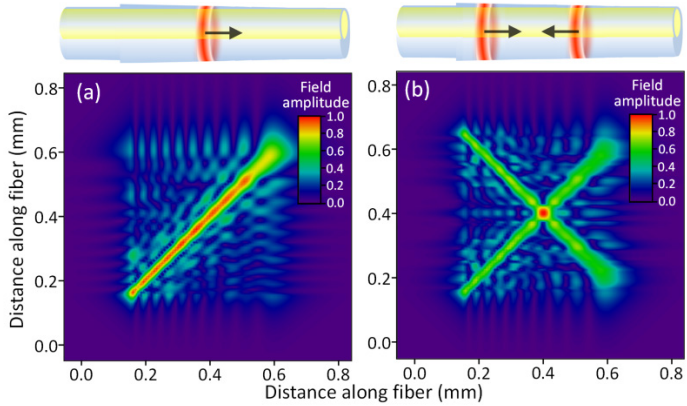

Fig. 3. (a) – The distribution of the driving field localized at the axial position defined by the vertical axis coordinate as a function of the horizontal axis coordinate. (b) – The distribution of the driving field localized at two axial positions defined by the vertical coordinate as a function of the horizontal axis coordinate. In both cases the input-output microfiber is positioned as indicated in Fig. 2.

The characteristic spacing between eigenvalues of the SNAP resonator considered has a relatively small value ~ 20 ps. This value grows with decreasing of the resonator length as $L^{-2}$ and is inverse proportional to the capillary radius $r_0$. For example, a two time shorter resonator has the characteristic resonance spacing ~ 0.1 nm. Notice that the available axial length for sensing and manipulation of microfluids can be increased by cascading the structures similar to that considers above.

In conclusion, it is shown here that the microfluidic processes can be detected and manipulated by monitoring the spectrum and varying the evanescent field of SNAP resonators in space and time. Tuning the input spectral distribution within a relatively narrow bandwidth of less than 1 nm allows one to program the axial distribution of the evanescent field and thereby manipulate the microfluidic components. The developed theory of sensing and manipulation of microparticles can be straightforwardly generalized to the case of a more complex distribution of refractive index and applied to investigation of complex spatial and temporal processes in the vicinity of the capillary surface. Similar theory is applicable to the investigation of media adjacent to the *outside surface* of the SNAP fiber (Fig. 1(a)). This is related to the processes in both liquids and gases. In particular, application of the developed approach to sensing and manipulation of clouds of cold atoms both outside [17-19] and inside [20, 21] the SNAP structure is of special interest. The practical realization of the proposed resonant slow light microfluidic sensor and manipulator can significantly advance the multi-disciplinary applications of optofluidics.

### References

1. T. M. Squires and S. R. Quake, "Microfluidics: Fluid physics at the nanoliter scale," Rev. Mod. Phys. **77**, 977 (2005).
2. G. M. Whitesides, "The origins and the future of microfluidics," Nature **442**, 368 (2006).
3. D. Psaltis, S. R. Quake and C. Yang, "Developing optofluidic technology through the fusion of microfluidics and optics," Nature **442**, 381 (2006).
4. X. Fan and I. M. White, "Optofluidic microsystems for chemical and biological analysis," Nat. Photon. **5**, 591, (2011).
5. D. Erickson, X. Serey, Y.-F. Chen, and S. Mandal, "Nanomanipulation using near field photonics," Lab Chip, **11**, 995 (2011).
6. F. Vollmer and L. Yang, "Label-free detection with high-Q microcavities: a review of biosensing mechanisms for integrated devices," Nanophotonics, **1**, 267–291 (2012).
7. S. Lin and K. B. Crozier, "An integrated microparticle sorting system based on near-field optical forces and a structural perturbation," Opt. Express **20**, 3367-3374 (2012).
8. C. Renaut, B. Cluzel, J. Dellinger, L. Lalouat, E. Picard, D. Peyrade, E. Hadji and F. de Fornel, "On chip shapeable optical tweezers," Scientific Reports **3**, 2290 (2013).
9. M. Soltani, J. Lin, R. A. Forties, J. T. Inman, S. N. Saraf, R. M. Fulbright, M. Lipson, and M. D. Wang, "Nanophotonic trapping for precise manipulation of biomolecular arrays," Nature Nanotechnology **9**, 448–452 (2014).
10. M. Sumetsky, D. J. DiGiovanni, Y. Dulashko, et al., "Surface nanoscale axial photonics: robust fabrication of high-quality-factor microresonators," Optics Lett. **36**, 4824 (2011).
11. M. Sumetsky, "Theory of SNAP devices: basic equations and comparison with the experiment," Optics Express **20**, 22537 (2012).
12. M. Sumetsky and Y. Dulashko, "SNAP: Fabrication of long coupled microresonator chains with sub-angstrom precision," Optics Express **20**, 27896 (2012).

13. M. Sumetsky, "Delay of light in an optical bottle resonator with nanoscale radius variation: dispersionless, broadband, and low loss", Phys. Rev. Lett. **111**, 163901 (2013).
14. I. M. White, H. Oveys, and X. Fan, "Liquid Core Optical Ring Resonator Sensors," Opt. Lett. **31**, 1319-1321 (2006).
15. X. Fan, I. M. White, H. Zhu, J. D. Suter, and H. Oveys, "Overview of novel integrated optical ring resonator bio/chemical sensors," Proc. SPIE **6452**, 6452M, 1-20 (2007).
16. Yu.N. Demkov, V.N. Ostrovskii, *Zero-Range Potentials and their Applications in Atomic Physics,* Plenum, New York, 1988.
17. J. R. Buck and H. J. Kimble, "Optimal sizes of dielectric microspheres for cavity QED with strong coupling," Phys. Rev. A **67**, 033806 (2003).
18. Y. Louyer, D. Meschede, and A. Rauschenbeutel, "Tunable whispering-gallery-mode resonators for cavity quantum electrodynamics," Phys. Rev. A **72**, 031801 (2005).
19. D. O'Shea, C. Junge, J. Volz, and A. Rauschenbeutel, "Fiber-Optical Switch Controlled by a Single Atom", Phys. Rev. Lett. **111**, 193601 (2013).
20. M. Bajcsy, S. Hofferberth, T. Peyronel, V. Balic, Q. Liang, A. S. Zibrov, V. Vuletic, and M. D. Lukin, "Laser-cooled atoms inside a hollow-core photonic-crystal fiber," Phys. Rev. A **83**, 063830 (2011).
21. J. A. Pechkis and F. K. Fatemi, "Cold atom guidance in a capillary using blue-detuned, hollow optical modes," Opt. Express **20**, 13409-13418 (2012).